\documentclass[%
 reprint,
superscriptaddress,
longbibliography,
 amsmath,amssymb,
 nobalancelastpage,
prb,
]{revtex4-1}
\usepackage[romanian,english]{babel}
\usepackage{combelow}
\usepackage{color}

\useshorthands{'}
\defineshorthand{'t}{\cb{t}}
\defineshorthand{'S}{\cb{S}}
\defineshorthand{'T}{\cb{T}}

\usepackage{graphicx}% Include figure files
\usepackage{dcolumn}% Align table columns on decimal point

\usepackage{bm}% bold math
\usepackage{amsmath}
\begin{document}

\preprint{APS/123-QED}

\title{Nature of the magnetic stripes in fully oxygenated La$_{2}$CuO$_{4+y}$}% Force line breaks with \\

\author{Ana-Elena 'Tu'tueanu}
 \affiliation{Nanoscience Center, Niels Bohr Institute, University of Copenhagen, 2100 Copenhagen, Denmark} 
 \affiliation{Institut Laue-Langevin, 38042 Grenoble, France}
 \email{tutueanu@ill.fr}
\author{Henrik Jacobsen}
\affiliation{Nanoscience Center, Niels Bohr Institute, University of Copenhagen, 2100 Copenhagen, Denmark}
\affiliation{%
Department of Physics, University of Oxford, OX1 3PU, UK
}%
\affiliation{Paul Scherrer Institute, Laboratory for Neutron Scattering and Imaging, 5232 Villigen, Switzerland}

\author{Pia Jensen Ray}
\affiliation{Nanoscience Center, Niels Bohr Institute, University of Copenhagen, 2100 Copenhagen, Denmark}

\author{Sonja Holm-Dahlin}
\affiliation{Nanoscience Center, Niels Bohr Institute, University of Copenhagen, 2100 Copenhagen, Denmark}

\author{Monica-Elisabeta L\u{a}c\u{a}tu\cb{s}u}
\affiliation{Nanoscience Center, Niels Bohr Institute, University of Copenhagen, 2100 Copenhagen, Denmark}
\affiliation{Paul Scherrer Institute, Laboratory for Neutron Scattering and Imaging, 5232 Villigen, Switzerland}

\author{Tim Birger Tejsner}
\affiliation{Nanoscience Center, Niels Bohr Institute, University of Copenhagen, 2100 Copenhagen, Denmark}
\affiliation{Institut Laue-Langevin, 38042 Grenoble, France}

\author{Jean-Claude Grivel}
\affiliation{Department of Energy Conversion, Technical University of Denmark, 4000 Roskilde, Denmark}

%\author{Niels Hessel Andersen}
%\affiliation{Department of Physics, Technical University of Denmark, 2800 Kgs.\ Lyngby, Denmark}

\author{Wolfgang Schmidt}
\affiliation{Forschungszentrum J{\"u}lich GmbH, J{\"u}lich Centre for Neutron Science at ILL, 38042 Grenoble, France}%

\author{Rasmus Toft-Petersen}
\affiliation{Helmholtz Center Berlin for Materials and Energy, 14109 Berlin, Germany}%
\affiliation{Department of Physics, Technical University of Denmark, 2800 Kgs.\ Lyngby, Denmark}

\author{Paul Steffens}
\affiliation{Institut Laue-Langevin, 38042 Grenoble, France}%

\author{Martin Boehm}
\affiliation{Institut Laue-Langevin, 38042 Grenoble, France}%

\author{Barrett Wells}
\affiliation{Department of Physics and Institute of Materials Science, University of Connecticut, 06269 Connecticut, USA}

\author{Linda Udby}
\affiliation{Nanoscience Center, Niels Bohr Institute, University of Copenhagen, 2100 Copenhagen, Denmark}

\author{Kim Lefmann}%
\affiliation{%
Nanoscience Center, Niels Bohr Institute, University of Copenhagen, 2100 Copenhagen, Denmark}%

\author{Astrid Tranum R\o mer}
\affiliation{Nanoscience Center, Niels Bohr Institute, University of Copenhagen, 2100 Copenhagen, Denmark}
\email{ar@nbi.ku.dk}

\date{\today}% It is always \today, today,

\begin{abstract}
We present triple-axis neutron scattering studies of static and dynamic magnetic stripes in an optimally oxygen-doped cuprate superconductor, La$_{2}$CuO$_{4+y}$, which exhibits a clean superconducting transition at $T_{\rm c}=42$~K. Polarization analysis reveals that the magnetic stripe structure is equally represented along both of the tetragonal crystal axes and that the fluctuating stripes display significant weight for in-plane as well as out-of-plane spin components. 
Both static magnetic order as well as low-energy fluctuations are fully developed in zero applied magnetic field and the low-energy spin fluctuations at $\hbar \omega = 0.3$-$10$~meV intensify upon cooling. We interpret this as an indication that superconductivity and low-energy spin fluctuations co-exist microscopically in spatial regions which are separated from domains with static magnetic order.

\end{abstract}

\maketitle

%\tableofcontents

\section{Introduction}
The interplay between magnetism and unconventional superconductivity (SC) remains controversial, inspiring much theoretical and experimental work in this field.~\cite{Vojta2009a,tranquada2013spins,Keimer2015,fradkin2015colloquium,agterberg2020physics} 
On the experimental side, one of the main pursuits is to establish key similarities in material properties across a range of existing superconductors, most of which differ in the details. For example, one such property is the magnetism, the fluctuations of which are thought to have an important role in the electron pairing mechanism.~\cite{ScalapinoReview02} Therefore, the magnetism of all known cuprate superconductors has been highly researched~\cite{julien2003magnetic,fujita2012progress,tranquada2013spins} since their discovery, but a general consensus on its role in the pairing mechanism is still to be reached.  

In this paper we address 
 La$_2$CuO$_{4+y}$ (LCOO) which belongs to the family of single-layer cuprate superconductors  La$_{2-x}$M$_x$CuO$_{4+y}$, where the metal dopant M is either Ba (LBCO) or Sr (LSCO). In LCOO, however, the doping is provided solely by the excess oxygen ions. All La-based cuprates exhibit optimal superconductivity at doping values of around 0.15 holes per formula unit, producing quite similar superconducting transition temperatures ($T_{\rm c}$), in the range $32$-$42$~K~.\cite{hucker2012structural, wells1997incommensurate, Kofu09}
In addition, these systems display incommensurate magnetic order observed by neutron scattering at a quartet of incommensurate (IC) values of the scattering vector, (in orthorhombic notation:) ${\bf Q}_{\rm IC} = (1\pm\delta_h, \pm\delta_k, 0)$, where often $\delta_h \simeq \delta_k \simeq \delta=1/8$.~\cite{yamada1998doping, Lee1999a, Kimura2000} The commensurate structure ($\delta=1/8$) is consistent with a picture of antiferromagnetic "stripes" of period 8, as first observed in Nd-containing LSCO,  La$_{1.48}$Nd$_{0.4}$Sr$_{0.12}$CuO$_4$, by Tranquada {\em et al.}~\cite{Tranquada1995} and later in LBCO
~\cite{Fujita2004,Hucker2011a}, LCOO~\cite{Lee1999a} and LSCO.~\cite{Suzuki1998} 

The magnetic order, also referred to as static stripes or spin-density-wave order (SDW), is generally regarded as a distinct electronic phase competing with uniform $d$-wave superconductivity, since it is predominantly present in underdoped samples with reduced $T_{\rm c}$. This interpretation is also supported by the fact that the magnetic ordering is more robust and survives to higher temperatures ($T_\textnormal{N}$) around the anomalous $x = 1/8$ doping, where $T_{\rm c}$ is significantly suppressed.~\cite{julien2003magnetic,Suzuki1998,Kimura2000,Katano2000} Additional support for the competition scenario is the observation of enhancement of the elastic IC response upon application of an external magnetic field perpendicular to the CuO$_2$ planes, indicating that local suppression of the superconducting order in vortex cores goes hand-in-hand with a strengthening of the magnetic order.~\cite{Lake2002,Chang2009,Brian2011,Hucker2011a,Hucker2011b,Wen2012a,Wen2012b} 

The corresponding magnetic fluctuations, also known as dynamic stripes, have been observed and comprehensively studied, especially in LSCO samples, over a wide range of dopings. In the absence of magnetic order, optimally doped LSCO samples exhibit gapped magnetic excitations with a gap of $\Delta \sim 6$~meV in the SC phase.~\cite{lake1999spin, yamada1995direct} In contrast, for slightly lower dopings $x < 0.13$ ($T_{\rm c} \leq 30$~K), where static stripes are prominent, the presence of the superconducting phase is marked by the opening of an incomplete spin gap.~\cite{chang2007magnetic, hiraka2001spin,Kofu09,jacobsen2015neutron} 
In both the optimally and underdoped cases, an applied magnetic field has been shown to induce sub-gap states as seen by an increase in spectral weight of low-energy fluctuations detected by inelastic neutron scattering experiments.~\cite{Lake01,Chang2009,Tranquada2004}

A special point in the phase diagram is found at the anomalous $1/8$ doping where both LSCO and LBCO show very different behaviour compared to the neighboring doping regimes. This effect is particularly strong in LBCO, where bulk superconductivity is almost fully suppressed, with $T_{\rm N} \sim 40$~K and $T_{\rm c} = 5$~K,~\cite{Fujita2004,Tranquada2008,Savici2005} although two-dimensional superconducting correlations still exist up to 40~K \cite{Li2007,Tranquada2008}. LBCO also displays pronounced  charge stripes with a periodicity of four lattice spacings, \cite{Tranquada2008} possibly related to a structural phase transition between the common cuprate low-temperature orthorhombic (LTO) phase and a low-temperature tetragonal (LTT) phase at a temperature close to the onset of charge order.~\cite{Fujita2004,Tranquada2008} The stripe order is very pronounced in LBCO, spanning a doping range of $0.095 < x < 0.155$.~\cite{Arai2003,Vojta2009a,hucker2012structural}
Charge stripes have also been found in LSCO~\cite{christensen2014bulk, croft2014charge, thampy2014rotated} 
and LCOO~\cite{ZhangWells2018}, both with hole concentration $\sim 1/8$.

For both LSCO and LBCO in the 1/8 phase, magnetic fluctuations display a small, partial gap  of the order 0.5~meV ascribed to spin anisotropy. Both gap and fluctuation intensities are insensitive to an external magnetic field.~\cite{Roemer13,Wen08} The absence of a full spin gap in LBCO has previously been interpreted as a deviation from uniform $d$-wave superconductivity towards a modulated phase of coexisting superconductivity and low-energy magnetic fluctuations in the form of a pair-density wave (PDW).~\cite{Xu2014} The magnetic fluctuations continue up to about 200~meV in an "hourglass" shape common for most cuprates.~\cite{tranquada2004quantum,NBC_hourglass_2004,Vojta2009a,fujita2012progress} 

In LCOO, there is a strong tendency towards electronic phase separation. \cite{Mohottala2006} 
Muon spin rotation experiments have been critical in demonstrating the spatial separation of the magnetic and superconducting phases.~\cite{Mohottala2006,Savici2002} Notable tools for characterizing each phase include neutron scattering,\cite{Savici2002,khaykovich2002enhancement,vaknin1987antiferromagnetism}, flux-pinning studies,\cite{mohottala2008flux} and resonant x-ray scattering.\cite{ZhangWells2018} The intercalant oxygen ions take up interstitial sites separated by a regular number of unit cell layers in a process known as staging. The separation between intercalant layers, the stage number, is a rough measure of the overall doping. For instance, samples with $y \sim 0.06$ are known to exhibit $T_{\rm c} = 32$~K and a stage-6 structure. Similarly, stage-4 samples have $T_{\rm c} = 42$~K and periodicity 4. 
%For example, certain values of the O-doping, $y$, will lead to the sample separating into phases of different doping and correspondingly different ordering temperatures. The only observed values are $T_{\rm c} = 20$~K, $32$~K, and $42$~K.~\cite{wells1997incommensurate,Mohottala2006} These precise doping levels are also accompanied by specific modulations of the intercalated oxygen atoms. For instance, samples with $y \sim 0.06$ are known to exhibit $T_{\rm c} = 32$~K and a so-called stage-6 structure with a periodicity 6 of the tilting of oxygen octahedra of interstitial oxygen layers (an ordered layer of interstitial oxygen appears at every 6th CuO$_2$ layer). Similarly, stage-4 samples have $T_{\rm c} = 42$~K and periodicity 4. Even for fully oxygenated samples ($T_{\rm c} = 42$~K everywhere), muon spin rotation experiments show a separation into a superconducting and a magnetic (striped) phase with almost equal volume fractions.~\cite{Mohottala2006,Savici2002} 

Neutron diffraction has shown strong static spin stripe signals in most LCOO samples, with a magnetic ordering temperature $T_\textnormal{N} \approx$ $T_{\rm c}$.~\cite{Mohottala2006} Early work on LCOO found that an applied magnetic field induces a significant enhancement of the neutron signal from the static stripes,\cite{Lee1999a,khaykovich2002enhancement,khaykovich2003effect} an observation which is similar to LSCO,\cite{Roemer13} but unlike LBCO.
~\cite{Dunsiger2008} Our sample, however, does not show a field-enhanced static signal and this feature thus appears to be a reflection of the extent of the static magnetic phase present in a particular sample. The dynamic stripes in LCOO with $T_{\rm c} = 32$~K~\cite{wells1997incommensurate} and $T_{\rm c} = 42$~K (stage-4)~\cite{Lee1999a} were previously addressed in neutron scattering studies. In Ref.~\onlinecite{wells1997incommensurate}, the IC signal from dynamic stripes at $\hbar\omega = 2$-$4$~meV was observed both above and below $T_{\rm c}$.
%Neutron diffraction has shown strong static spin stripe signals in LCOO, with a magnetic ordering temperature $T_\textnormal{N} \approx$ $T_{\rm c}$, which is not accompanied by a decrease in $T_{\rm c}$.~\cite{Mohottala2006} Previous work on LCOO found that an applied magnetic field induces a significant enhancement of the neutron signal from the static stripes.~\cite{Lee1999a, khaykovich2002enhancement, khaykovich2003effect}, an observation which is similar to LSCO,\cite{Roemer13} but unlike LBCO. \cite{Dunsiger2008} Our sample, however, does not show a field-enhanced static signal. Furthermore, the dynamic stripes in LCOO with $T_{\rm c}=32$~K (stage-6),~\cite{wells1997incommensurate} and $T_{\rm c} =42$~K (stage-4)~\cite{Lee1999a} were previously addressed in neutron scattering studies. 
In Ref.~\onlinecite{wells1997incommensurate},  the IC signal from dynamic stripes at $\hbar\omega = 2$-$4$~meV was observed both above and below $T_{\rm c}$.

Here we revisit the low-energy spectrum of LCOO with the objective 
%to decode the complex interplay between magnetism and superconductivity %in this family of materials,
%and 
to test the universality of magnetic order and fluctuations, compared to its cousins LBCO and LSCO, including the response to different parameter variations, such as magnetic field and temperature. LCOO is remarkably different from LBCO, and to some extent to LSCO, in that the appearance of static stripes does not suppress $T_{\rm c}$. It is therefore relevant to search for underlying similarities in other parts of their magnetic spectra. For this reason, we here present the results of a comprehensive study of static and dynamic spin stripes in LCOO. 

Our main findings are the absence of a spin gap in the low-energy fluctuation spectrum in the superconducting state and the absence of a significant magnetic field effect both in the elastic and inelastic channels. Furthermore, our $xyz$ polarization analysis reveals that the static stripes do not have any preferential direction between the two in-plane tetragonal $a_{\rm T}$ and $b_{\rm T}$ axes in this system. %This could suggest an orthogonal stripe arrangement similar to the proposed PDW order in LBCO,~\cite{agterberg2020physics} but other probes, which are sensitive to the interlayer phase coherence are necessary for a conclusive test of this hypothesis.

\section{Experimental method}
% \label{section_exp}

\begin{figure}[b!]
\centering
\includegraphics[width=0.49\textwidth]{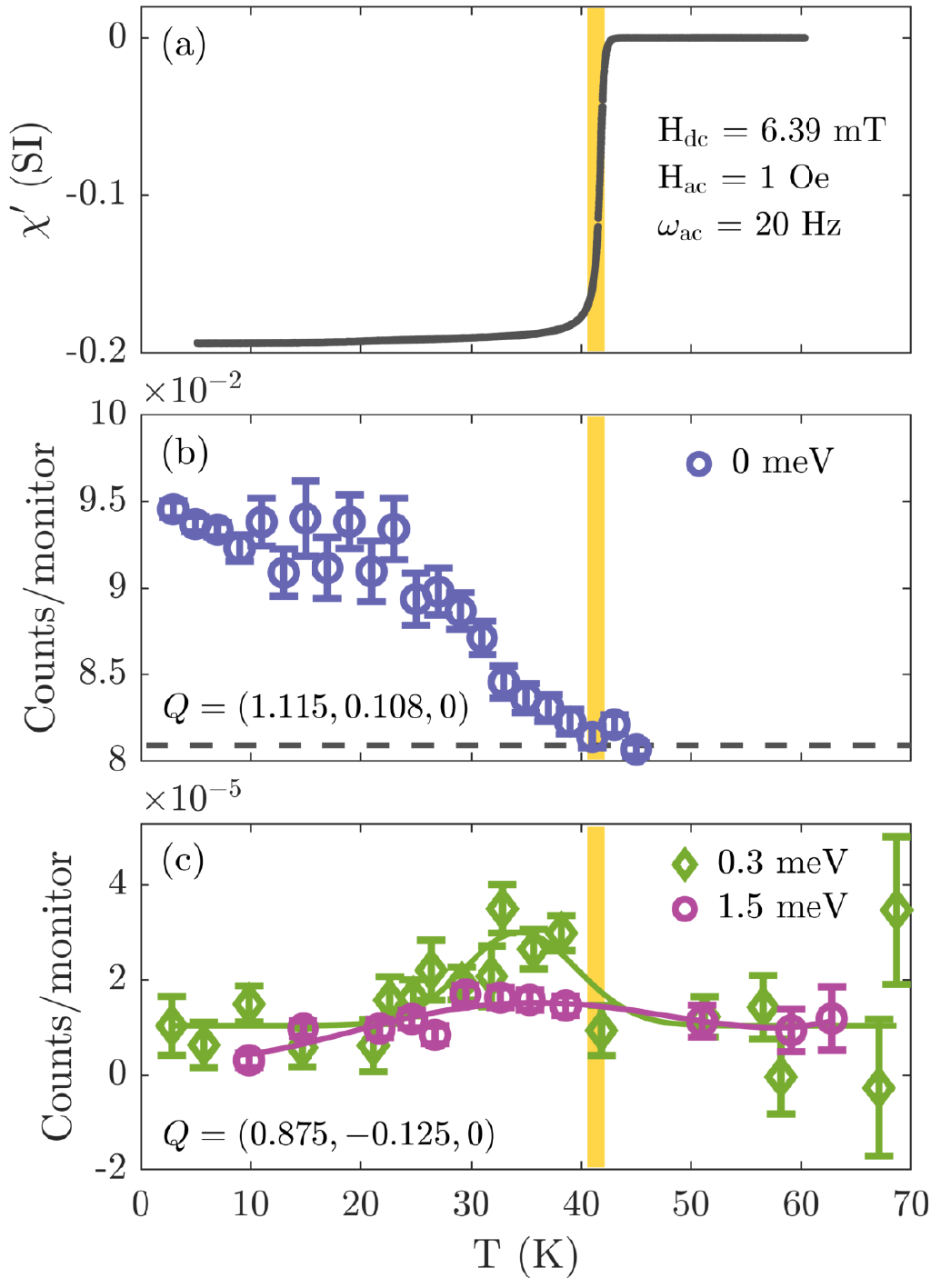}
\caption{\textbf{a)} The real part of AC magnetic susceptibility measurements showing the sharp transition to a diamagnetic state upon cooling below the superconducting critical temperature. $T_{\rm c}$ is indicated by the yellow stripe which spans a temperature range $41.3\pm 0.8$~K. \textbf{b)} Temperature dependence of magnetic order measured as 1-point scans on top of one of the IC peaks $(1.115, 0.108, 0)$. The incoherent background has not been subtracted. The dashed line indicates the background level extracted from an energy scan taken away from the incommensurate peak position, at $(1.1, -0.24, 0)$. \textbf{c)} Temperature dependence of magnetic fluctuations with energy $0.3$~meV and $1.5$~meV. Data were measured as 3-points scans and the intensity is obtained as the difference between the point on top of the peak, at $(0.875,−0.125, 0)$, and the average of the 2 background points. }
\label{fig:susceptibility and temp dependence elastic}
\end{figure}

The samples used throughout this study were prepared by growing stoichiometric crystals of La$_2$CuO$_4$ (LCO) at the Technical University of Denmark in an optical image furnace using the traveling solvent float zone technique.~\cite{CrystalGrowth} After annealing and X-ray characterization, chosen crystals were super-oxygenated in an aqueous bath at the University of Connecticut. The resulting LCOO crystals were cut into pieces of $3$-$4$~g each, suitable for neutron scattering experiments. Smaller pieces of the samples were used for magnetic susceptibility measurements which revealed a single superconducting transition temperature of $41.3 \pm 0.8$~K (midpoint), indicating the presence of a unique superconducting phase,~\cite{Mohottala2006} see Fig.~\ref{fig:susceptibility and temp dependence elastic}(a).

The data discussed in this paper were acquired during several neutron scattering experiments carried out on the triple-axis spectrometers FLEXX,~\cite{habicht2015upgraded, Le2013} at the Helmholtz Center Berlin (HZB), IN12,~\cite{Schmalzl2016,IN12_data} a J{\"u}lich Centre for Neutron Science (JCNS) instrument outstationed at the Institut Laue-Langevin (ILL) and ThALES,~\cite{boehm2015thales, ThALES_data, ThALES_data_pol} at the ILL. Results from one of the ThALES experiments were previously reported in another context in Ref.~\onlinecite{Jacobsen2018}. Each experiment followed the evolution of both the magnetic order and spin fluctuations under different conditions such as: varying temperature (IN12 and the first Thales experiment, Th1, see Figs.~\ref{fig:susceptibility and temp dependence elastic} and~\ref{fig:raw data}), applied magnetic field (FLEXX, see Fig.~\ref{fig:susceptibility field depednence})  and in polarization analysis configuration in order to determine the in-plane orientation of the magnetic moments (second Thales experiment, Th2, see Fig.~\ref{fig:polarization}).

All instruments used for this project employ a velocity selector on the incident beam, before the monochromator, in order to remove second-order contamination. In addition, a cooled Be-filter between sample and analyzer was used in the second part of the IN12 experiment and in the Th1 experiment to further reduce background. The instrument set-up on IN12 contained vertically and horizontally focusing monochromators, leading to a relaxed in-plane $Q$-resolution of about $0.02$~$\textnormal{\AA}^{-1}$ (for $k_{\rm f} =  1.5~$\AA$^{-1}$) and even broader resolution out of the scattering plane, where the stripe signal from cuprates is nearly constant.~\cite{Roemer2015,Lee1999a} In the Th1 experiment only vertical focusing was employed, leading to an enhanced in-plane momentum resolution by a factor 2 ($0.01$~$\textnormal{\AA}^{-1}$) and relatively loose resolution along the $c$-direction. For the polarization analysis (Th2) the instrument configuration contained Heusler (111) monochromator and analyzer as well as an orange cryostat placed inside a cryopad module~\cite{tasset1989zero} for accurate control of the spin polarization. 

The sample environments used were orange cryostats for the experiments performed at the ILL and a 15~T cryomagnet for the experiment performed at the HZB. For all experiments, the sample was aligned in the $a$-$b$ plane, enabling access to the $\mathbf{Q} = (h, k, 0)$ scattering plane. Throughout the paper, the orthorhombic notation is used where the size of the unit cell is $a_{\rm o} = 5.33$~$\textnormal{\AA}$, $b_{\rm o} = 5.40$~$\textnormal{\AA}$, $c_{\rm o} = 13.20$~$\textnormal{\AA}$ and the antiferromagnetic reflection is found along $\mathbf{Q}_{\textnormal{AFM}} = (1, 0, 0)$. However, due to twinning commonly present in these samples, each scattering point is a superposition of $(h,0,0)$ and $(0,k,0)$ reflections,~\cite{wells1996intercalation} meaning that antiferromagnetic scattering is also observed along $\mathbf{Q}_{\textnormal{AFM}} = (0, 1, 0)$ without deviation of the spins from the orthorhombic $b$-axis.~\cite{vaknin1987antiferromagnetism} It should be mentioned that we have observed elastic intensity at the AFM reflection and confirmed its magnetic origin by polarized neutron scattering. In a homogeneously doped sample, AFM order would not be expected. However, in oxygen doped samples phase separation into oxygen-rich and oxygen-poor regions~\cite{wells1997incommensurate} might result in a remnant AFM order in small (presumably undoped) parts of the crystal. From this point forward, when referring to magnetic order we allude to the incommensurate order generated by the spin stripe arrangement. 

During all experiments the sample was cooled slowly ($1$~K/min) in the temperature range 300~K to 100~K, in order to prevent unwanted effects arising from quenched oxygen disorder.~\cite{lee2004neutron, Roemer2015} Measurements were performed in the SC phase at $2$~K and in the normal phase at $45$~K. In the experimental setup, we resolve all four IC magnetic peaks at $\mathbf{Q}_{\textnormal{IC}} = (1 \pm 0.125, 0\pm \delta, 0)$ which are resolution limited.
The data were acquired through scanning over one or two of the four peaks with energy transfers between $0$ and $10$~meV. We furthermore followed the temperature dependence of the intensity of the peaks at $0$~meV, $0.3$~meV and $1.5$~meV using one- and three-point measurements. The same crystals in different co-alignment combinations have been used for all experiments. The total masses of the samples used in each experiment were as follows: FLEXX $8.84$~g, IN12 $15.79$~g and ThALES $3.44$~g (one single crystal). In the FLEXX and IN12 experiments the crystals were co-aligned to within 1.3$^\circ$ and 2$^\circ$, respectively. For this particular study, where magnetic scattering is in focus, the use of greater sample mass allowed us to overcome the counting time restrictions imposed by the small cross section of inelastic magnetic scattering. With an increased neutron flux at the sample position, such as is the case at ThALES, a lower sample mass can provide better signal to noise ratio, since the size of the beam hitting both the cryostat and the sample can be significantly reduced.

Another manner of improving the signal to noise ratio is to tune the size of the resolution ellipsoid by varying the wave vector of the scattered neutrons, $k_{\rm f}$. By using a smaller value of $k_{\rm f}$, we obtain a better energy resolution, which is preferable when studying low energy excitations, where tails from the elastic scattering can greatly contribute to the inelastic background signal. For this reason, we chose to employ different $k_{\rm f}$ values ($k_{\rm f} = 1.1, 1.2, 1.5, 1.55~$\AA$^{-1}$) when measuring excitations of different energies across the two experiments (IN12 and Th1) which are combined in Figure~\ref{fig:raw data}(c). 
  
To account for the difference in experimental setup, we normalize the measured raw counts in three steps. First we normalize to the monitor count, as shown in Fig.~\ref{fig:raw data} (a, b), which is inversely proportional to the incoming wave vector ($k_{\textnormal{i}}$). Taking into account that the scattering cross section is also proportional to $1/k_{\rm i}$, we obtain a scattering signal per monitor, where the dependence of $k_{\textnormal{i}}$ is normalized out.\cite{Shiranebook} In order to obtain the susceptibility, $\chi''$ in units of $\mu_B^2$eV$^{-1}$Cu$^{-1}$ similar to the procedure performed in Ref.~\onlinecite{xu2013absolute}, we measured a low energy acoustic phonon at $\mathbf{Q}=(2,0,0)$, for $k_\textnormal{f}=1.5$~\AA{}$^{-1}$ at IN12 and for $k_\textnormal{f}=1.55$~\AA{}$^{-1}$ at Thales (Th1). Lastly, for the IN12 data taken at other outgoing wave vectors, we corrected for the change in resolution volume~\cite{Shiranebook} using a factor proportional to the ratio between the resolution volume ($V_\textnormal{f}$) corresponding to $k_\textnormal{f}=1.5$~\AA{}$^{-1}$ (at which the phonon was also measured) and the ones corresponding to the different $k_\textnormal{f}^\prime$ values at which each data set was taken ($V_\textnormal{f}^\prime$). We have used the following simplified ratio where we assumed constant reflectivity of the analyzer for the various outgoing wave vectors:  

\begin{equation}
    \frac{V_\textnormal{f}}{V_\textnormal{f}^\prime} = \frac{k_\textnormal{f}^3}{\tan(\mathrm{\theta}_A)} \frac{\tan(\mathrm{\theta}_A^\prime)}{k_\textnormal{f}^{\prime 3}}, 
\end{equation}
where $\rm \mathrm{\theta}_A$ and $\rm \mathrm{\theta}'_A$ are the Bragg angles of the analyzer for the two $k_\textnormal{f}$ values. 

We note that %because of the combination of a number of different experiments, 
in Fig.~\ref{fig:raw data} we present the dynamic susceptibility at the peak position, $\chi''(\mathbf{Q}_\text{peak},\omega)$, rather than the average over the Brillouin Zone to obtain $\chi''(\omega)$. We use this approach because averaging over the Brillouin zone requires knowledge of the signal along all directions in reciprocal space and we did the measurements only along one cut. However, in the Supplementary Material we provide an example of $\mathbf{Q}$-integrated susceptibility for the Thales data where we assume incommensurate peaks of the same width along all directions, and compare the result of this analysis with LBCO measurements of Ref.~\onlinecite{Xu2014}. In addition, the Appendix includes a detailed description of all the steps of the normalisation process.
The appeal of absolute normalisation is the possibility to directly compare measurements performed with different sample mass on various instruments as well as a quantitative comparison between the magnetic susceptibilities of different compounds.

 We have used two fitting routines. For the temperature dependence (Fig.~\ref{fig:raw data}) we implemented a so-called Sato-Maki function, first proposed as a model of describing antiferromagnetic correlations in chromium and its alloys.~\cite{SatoMaki} This was later comprehensively explained and used by Aeppli {\em et al.}\ to fit neutron scattering measurements of magnetic fluctuations in optimally doped LSCO.~\cite{aeppli1997nearly} While Aeppli {\em et al.}\ used a tetragonal notation, we have converted the formalism to the orthorhombic structure. The associated lattice parameters $a_{\rm o} = b_{\rm o} = 5.3$~$\textnormal{\AA}$ have been used, since these are too close to distinguish from one another given the resolution of our inelastic neutron scattering experiments. The scattering amplitude is thus fitted to the function
  \begin{equation}
     S(Q,\omega) = \frac{[n(\omega)+1]\chi^{\prime\prime}(\omega, T)\kappa^4(\omega, T)}{[\kappa^2(\omega, T)+R(Q)]^2},
 \end{equation}
 where 
 \begin{equation}
     R(Q) = \frac{4[(Q-Q_{\rm AFM})^2-\delta^2]^2}{(2a_o\delta)^2}.
 \end{equation}
 Here $(n(\omega)+1)$ is the thermal population factor for down-scattering (neutron energy loss), $\chi^{\prime\prime}(\omega, T)$ is the imaginary part of the dynamic susceptibility at the peak position, $\kappa(\omega, T)$ is the peak width and $\delta$ is the incommensurability of the signal measured around the antiferromagnetic point ($\mathbf{Q}_{\rm AFM}$).
 Note that the possible differences in incommensurabilities $\delta_h$ and $\delta_k$ are small compared to the width of the resolution function~\cite{Jacobsen2018} and we simply set $\delta_h=\delta_k=\delta$ henceforth.
 This routine imposes equal amplitude, width, and incommensurability for the two peaks that are scanned over during the measurement. The incommensurability is defined as the distance between the peak center and $\mathbf{Q}_{\rm AFM}$=(1,0,0).
 
 To analyze the rest of the data we used a Gaussian model with constrained equal widths and incommensurability, because part of the data were obtained by scanning over a single peak (Fig.~\ref{fig:susceptibility field depednence}) or due to the appearance of additional parasitic scattering on top of one of the peaks (Fig.~\ref{fig:polarization}).

\section{Results}

The sample exhibits a single and clean superconducting transition at $T_{\rm c}=42$~K, as seen from the temperature dependence of the diamagnetic signal shown in Fig.~\ref{fig:susceptibility and temp dependence elastic}(a). This supports the presence of a single dominant staging structure which, given its critical temperature, was previously found to correspond to stage-4 samples.~\cite{wells1997incommensurate, wells1996intercalation} The temperature dependence of the magnetic order in Fig.~\ref{fig:susceptibility and temp dependence elastic}(b) unveils that the onset temperature of the elastic stripe signal coincides, within errors, with the onset of superconductivity, $T_{\rm c} \approx T_{\rm N}$.

The low-energy magnetic fluctuations have a much higher onset temperature than superconductivity ($T_{\rm onset}> 70$~K $\, >$ $T_{\rm c}$) and do not show any dramatic signatures at $T_\textnormal{N}$ or $T_{\rm c}$, see Fig.~\ref{fig:susceptibility and temp dependence elastic}(c). At the lowest energy $\hbar \omega=0.3$~meV, a broad increase in intensity is observed around $T=30$~K ($ <$ $T_{\rm c}$). A similar temperature dependence of fluctuations of slightly higher energy ($\hbar \omega=2$~meV) was observed by Y. Lee {\em et al.}~\cite{Lee1999a} in a stage-4 LCOO sample. These measurements showed an increase in intensity down to $\sim 30$~K, followed by a less pronounced suppression at lower temperatures. In stage-6 samples, some of us earlier reported~\cite{wells1997incommensurate} the same temperature behaviour of low energy magnetic fluctuations ($2$-$4$~meV). In that work, the peak in intensity happened to coincide, within errors, with the critical temperature $T_{\rm c} = 31$~K and to follow the same trend as underdoped ($x=0.12$) LSCO samples of similar $T_{\rm c}$.~\cite{Roemer13}
The present data show that the increase in intensity of the low-energy fluctuations at temperatures close to 30~K is likely a universal feature of LCOO and underdoped LSCO superconductors regardless of their corresponding superconducting critical temperatures.

\begin{figure}[h!]
\includegraphics[width=0.5\textwidth]{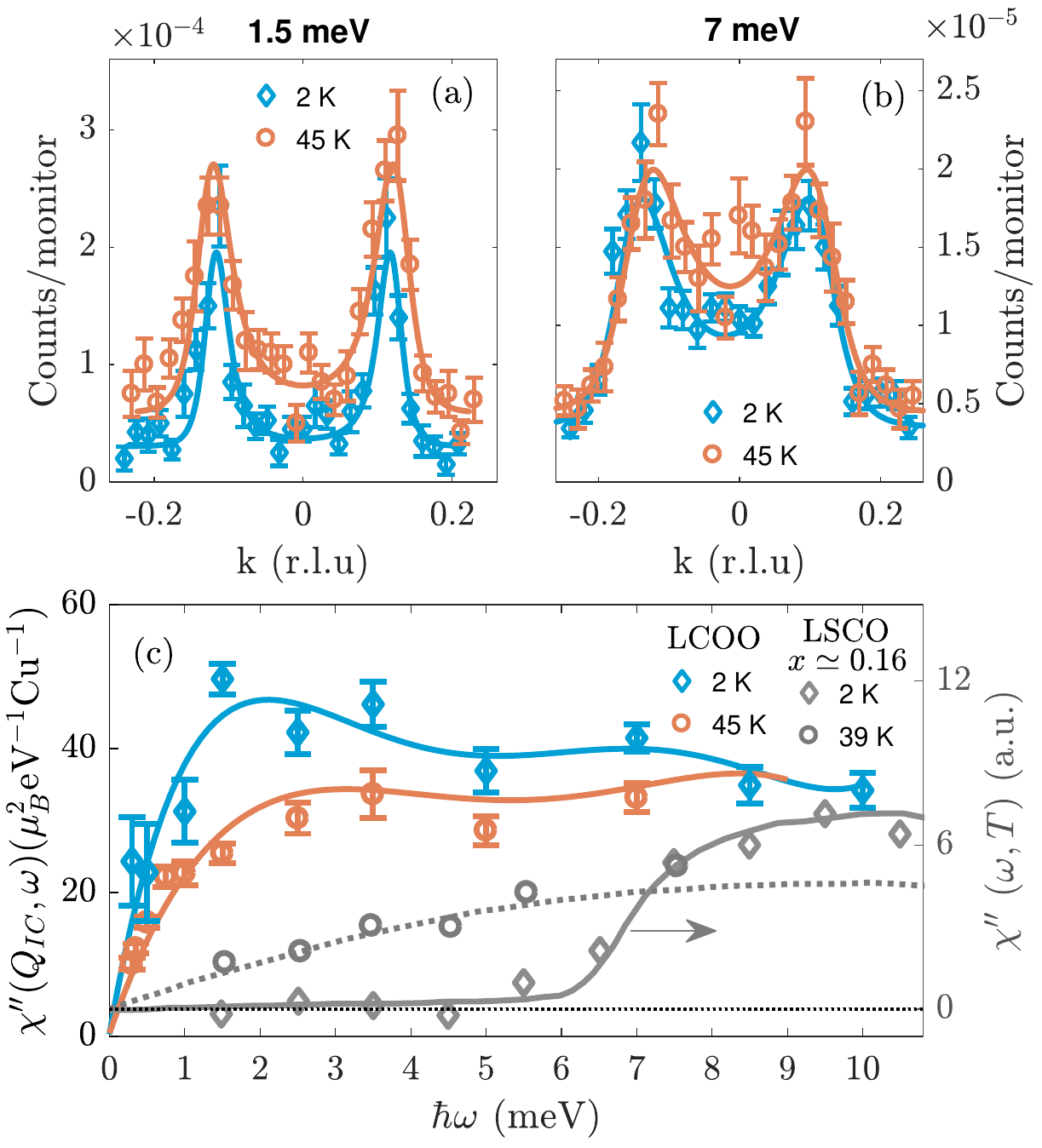}
\caption{\textbf{a)} and \textbf{b)} Representative inelastic scans through the incommensurate positions ($\pm \delta, 1-\delta, 0$) where $\delta = 0.125$, collected at $1.5$~meV and $7$~meV with fixed outgoing wavevectors $k_{\textnormal{f}} = 1.55$~$\textnormal{\AA}^{-1}$ and $k_{\textnormal{f}} = 1.5$~$\textnormal{\AA}^{-1}$, respectively. The data were collected at ThALES (\textbf{a}) and  IN12 (\textbf{b}) at two temperatures: blue diamonds show data in the superconducting state (2~K) and orange circles show data within the normal state (45~K). The solid lines are Sato-Maki function fits to the raw data as described in the text. \textbf{c)} Dynamic susceptibility $\chi''({\bf Q}_{\textnormal{IC}},\omega)$ measured inside ($2$~K - in blue) and outside ($45$~K - in orange) the superconducting dome. Solid lines are guides to the eye. Each data point represents an average of values obtained at different k$_{\textnormal{f}}$ values and on different instruments (ThALES and IN12). In grey symbols, we show for comparison the magnetic susceptibility measured inside and outside the superconducting state by B. Lake {\em et al.}~\cite{Lake01} on a optimally doped $x=0.163$ LSCO sample.}
\label{fig:raw data}
\end{figure}

We now address how cooling below $T_{\rm c}$ influences the low-energy spectrum in the energy range $\hbar \omega=0.3$-$10$~meV. Figures~\ref{fig:raw data}(a,b) show examples of representative scans taken at Thales and IN12, respectively, at temperatures both inside (2~K) and outside (45~K) the superconducting phase. The apparent increase in integrated intensity at the low energies as a function of increased temperature (Fig.~\ref{fig:raw data}(a)) is merely an effect of the Bose occupation factor. Fig.~\ref{fig:raw data}(c) depicts the imaginary part of the dynamical  spin susceptibility $\chi''({\bf Q}_{\textnormal{IC}},\omega)$,  expressed in units of ($\mu_B^2$eV$^{-1}$Cu$^{-1}$).~\cite{Shiranebook, xu2013absolute} The data demonstrate that spin fluctuations with energy $\hbar\omega > 4$~meV are rather insensitive to the onset of superconductivity. At lower energies, cooling below $T_{\rm c}$ leads to an increase in the spin susceptibility at ${\bf Q}_{\rm IC}$.
 This does not appear to be a direct consequence of the onset of superconductivity. The data exhibit no tendency towards a suppression of the low-energy magnetic spectrum in the superconducting phase and the sample thus shows no evidence of a spin gap, although a small partial gap at $\hbar\omega \sim 0.5$~meV cannot be excluded.

Our observations are in contrast to studies of optimally doped LSCO, with a similar high value of $T_{\rm c}$, where the superconducting transition is accompanied by the opening of a clean spin gap in the low-energy magnetic spectrum.~\cite{Kofu09,Lake01} Subsequently, an increase in temperature above $T_{\rm c}$ induces sub-gap states. To visualize this difference, we plot in Fig.~\ref{fig:raw data}(c) the data obtained in this sample together with the data of LSCO $x=0.163$ from Ref.~\onlinecite{Lake01}. On the other hand, the low-energy spectrum bear close resemblance to LBCO~\cite{Xu2014} and LSCO $x=0.12$~\cite{Roemer13} despite the higher $T_{\rm c}$ and the tendency towards phase separation not being observed in the latter compounds.

\begin{figure}[h!]
\includegraphics[width=0.45\textwidth]{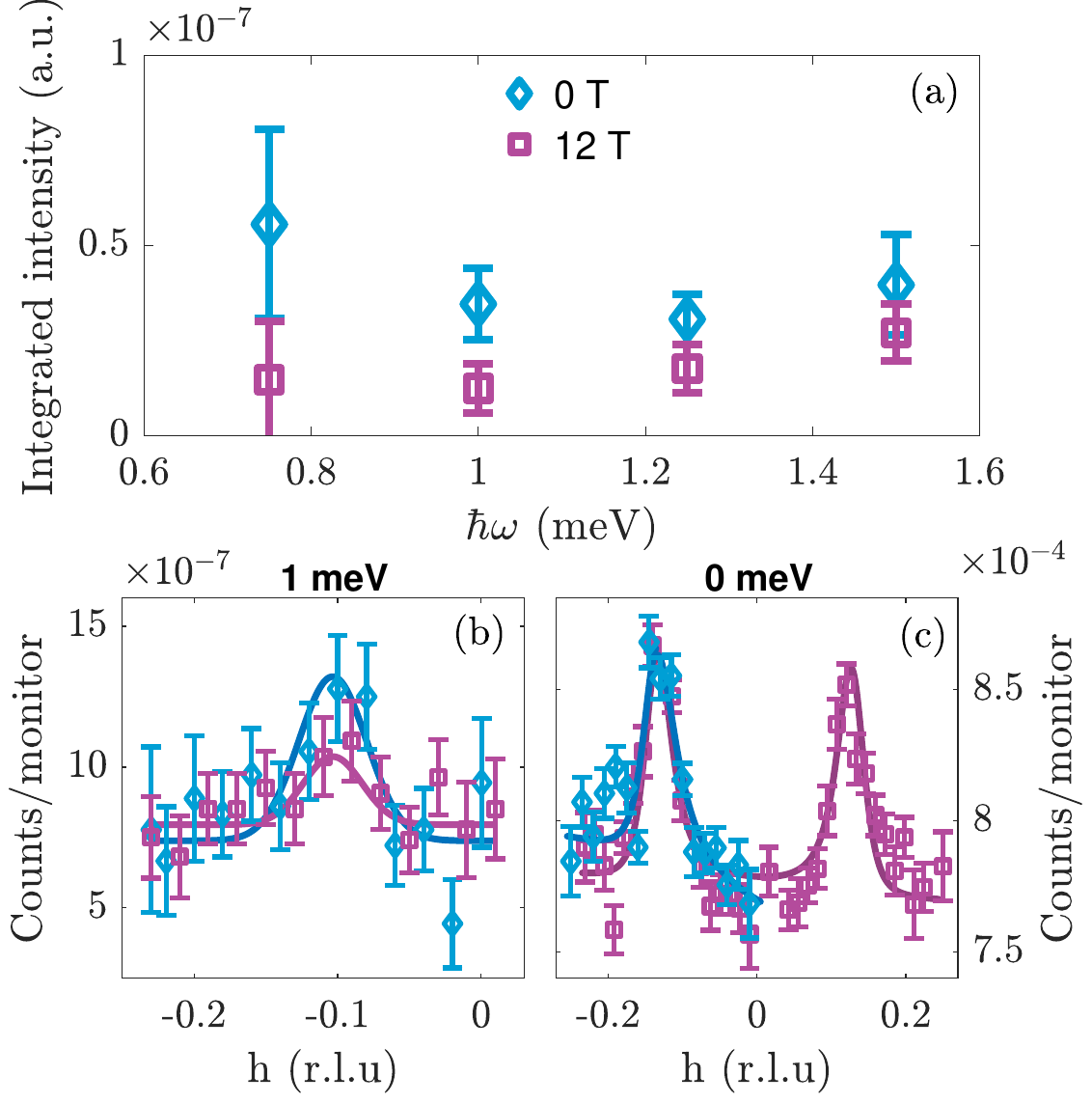}
\caption{\textbf{a)} Magnetic field dependence of low-energy spin excitations measured at $2$~K. Representative \textbf{b)} inelastic constant energy scan ($\hbar\omega = 1$~meV) and \textbf{c)} elastic scans collected in the superconducting phase ($2$~K) with and without applied magnetic field. The solid lines are Gaussian fits to the raw data.}
\label{fig:susceptibility field depednence}
\end{figure}

In order to gain a more comprehensive understanding of the magnetism of our oxygen-doped sample, we have also pursued the effect of an applied magnetic field. This experiment was performed on the FLEXX spectrometer, and since measurements were acquired mostly as scans over one of the incommensurate peaks, a single Gaussian function was used to fit the data. In addition, no Bose-scaling was applied because the temperature was kept constant and the energy difference produces a negligible effect in this regime. Along the $c$-axis, a $H=12$~T magnetic field was applied, which is much lower than the upper critical field $H_{c2} \sim 60$~T (see Ref.~\onlinecite{ando1999resistive}). The data shown in Fig.~\ref{fig:susceptibility field depednence} illustrate the lack of response of the elastic channel and a small decrease with field in the low-energy inelastic channel. The insensitivity of magnetic order to an applied magnetic field contradicts reports from the literature on LCOO. We will comment on this aspect in Section~\ref{discussion field effect}.

% ---- Polarization analysis: ---

To access the orientation of the magnetic moments in static and dynamic magnetic stripes, we employed polarized neutron scattering analysis with three different spin configurations of the incoming neutron beam. First, the neutron polarization was chosen parallel to the scattering vector {\bf Q}, defined as the $x$-direction in the following. In this set-up, we performed scans in both the spin flip (SF) and the non-spin-flip (NSF) configurations. Afterwards, only the SF channel was measured with the neutron spin aligned along the $y$ and $z$-direction, $z$ being out of the scattering plane. This $xyz$ polarization analysis allowed us to determine the contribution of different components of the magnetic order in the sample, as illustrated in Table~\ref{table:magnetization}.

\begin{figure}[h!]
\centering
\includegraphics[width=0.48\textwidth]{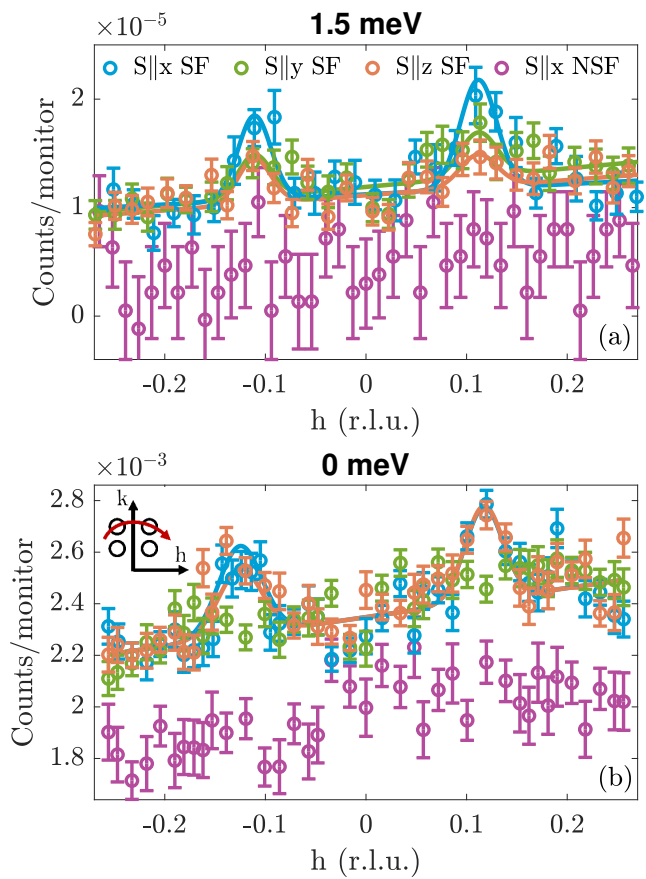}
\caption{\textbf{a)} Inelastic constant energy scan ($\hbar\omega = 1.5$~meV) and \textbf{b)} elastic scans collected in the superconducting phase (30~K and 5~K respectively) in spin-flip (SF), with 3 spin configurations, and non-spin-flip (NSF) mode. The scan direction is represented in the inset of subplot (b). In panel (a) the NSF data are displaced by a factor $-0.7\times 10^{-5}$ for better visualisation. $S$ denotes the incoming neutron beam spin direction. The solid lines are Gaussian fits to the raw data.}
\label{fig:polarization}
\end{figure}

In the elastic channel, Fig.~\ref{fig:polarization}(b), we observe a lack of intensity in the $S\parallel y$ spin flip channel and equal intensity in the other two SF channels. Compared to Table~\ref{table:magnetization}, this provides definite evidence that  $M_{zz}$ is (close to) zero and thus that the spin structure in the elastic channel resides in the $a$-$b$ plane. Previous reports,~\cite{vaknin1987antiferromagnetism, Lee1999a} found that the spin direction of the parent compound, which is along the orthorhombic $b$-axis, is preserved upon doping and we therefore assume that the spins are aligned along the $b$-axis in our sample. This assumption becomes important in the discussion presented in Section~\ref{sec:discussion_polarized}, where we attribute the IC signals to different twin domains and stripe orientations.

The inelastic polarized data of Fig.~\ref{fig:polarization}(a) shows scattering intensity in all spin channels. This means that there is an out-of-plane spin component to the scattering signal, which is expected both in the case of isotropic spin fluctuations, and in the limit of purely transverse fluctuations connected to the static SDW signal. In Section~\ref{sec:discussion_polarized} we discuss the polarized measurements in further detail.

\begin{table}[h]
\centering
\begin{tabular}{|c|c|c|}
  \hline
   Spin direction & NSF & SF  \\ \hline
  $S \parallel x$ & 0 & $M_{yy} + M_{zz}$ \\ \hline
  $S \parallel y$ & $M_{yy}$ & $M_{zz}$ \\ \hline
  $S \parallel z$ & $M_{zz}$ & $M_{yy}$ \\ \hline
\end{tabular}
  \caption{Expected components of the magnetic correlation function, $M$, measured in different spin configurations of the incoming neutron beam, $S$. In this table, $M_{\alpha \alpha}=\frac{1}{2\pi\hbar}\int dt e^{i\omega t} \langle M^\dagger_\alpha({\bf Q}_{\rm IC},0)  M_\alpha({\bf Q}_{\rm IC},t) \rangle$, and $M_\alpha({\bf Q}_{\rm IC},t)$ is the spin component ($\alpha=x,y,z$) at the incommensurate position at time $t$. We have assumed that there is no chiral contribution to the $S \parallel x$ channel. Contributions from incoherent scattering and nuclear coherent scattering are omitted for simplicity as these would give rise to the same background within all the spin channels. A more detailed overview of the formalism involved can be found e.g.  in Refs.~\onlinecite{jacobsen2018spin} and~\onlinecite{Boothroydbook}.}
  \label{table:magnetization}
  \end{table}

\begin{figure*}[t!]
\centering
\includegraphics[trim={0cm 19.3cm 9.8cm 0cm},clip,width=\textwidth]{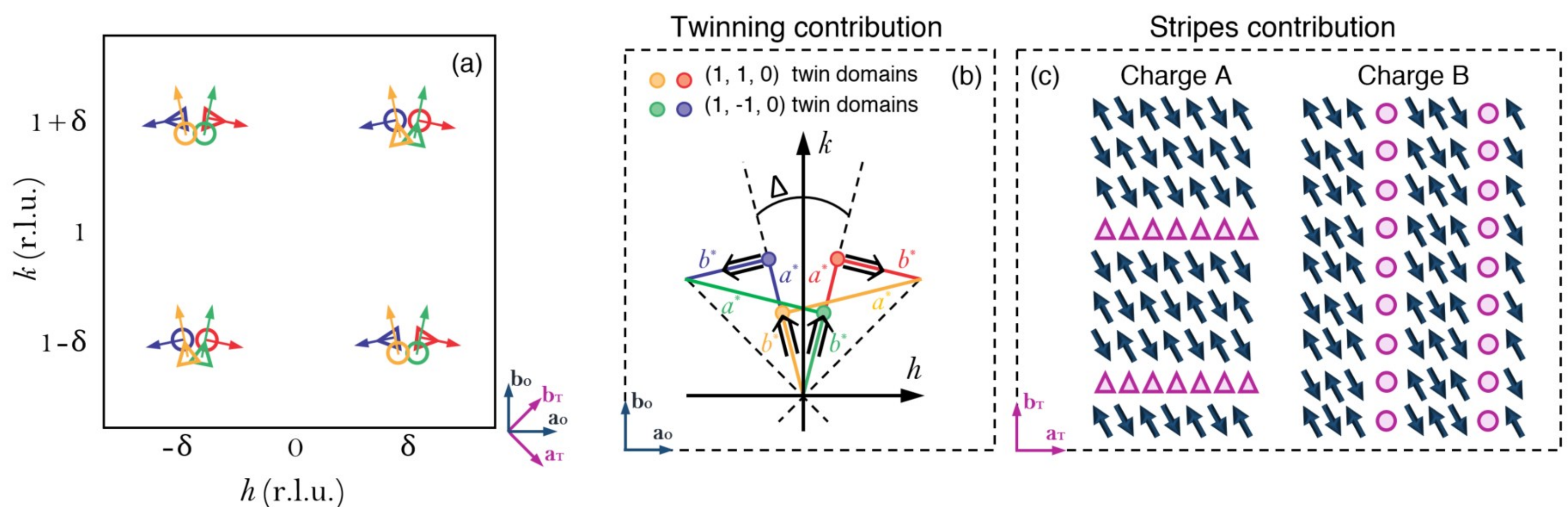}
\caption{\textbf{a)} A cartoon illustration of the neutron scattering signal generated by the incommensurate AFM spin structure including contributions from all four crystal twins. \textbf{b)} Twinning of a structural peak into two domains each composed of a pair of twins. The direction of the orthorhombic $b$-axis, which corresponds to the spin direction, is shown by a double-arrow for each twin peak. $\Delta$ is the angular separation between two twin reflections~\cite{braden1992characterization} and in our orthorhombic system has a value $\Delta = 90^\circ - 2 \arctan(\frac{a}{b})\sim 0.8^\circ$, a splitting too small to be resolved given our $Q$-resolution. $a^*$ and $b^*$ are the reciprocal lattice constants in orthorhombic notation. Image adapted from the Supplemental Material of Ref.~\onlinecite{Jacobsen2018}. \textbf{c)} The stripe pattern arising from charge stripes running along the tetragonal $a_\textrm{T}$-axis (Charge A) and the tetragonal $b_\textrm{T}$-axis (Charge B). The triangles and circles indicate how the charge orientations corresponds to the incommensurate peak structures in (a) for the local coordinate system of each twin type, which are rotated by 90 degrees with respect to each other.}
\label{fig:pattern}
\end{figure*}

\section{Discussion}
\label{section_dis}
\subsection{Stripe structure and twinning}
\label{sec:discussion_polarized}
To analyze the structure of the spin stripe modulation, we consider two arrangements of the charge stripes which are assumed to accompany the magnetic stripes as antiphase domain walls.~\cite{ZhangWells2018} The two possible arrangements of charge stripes along the tetragonal axes are depicted in Fig.~\ref{fig:pattern}(c). 
To access information about the structure of the spin stripe modulation 
%induced by the different arrangements of  charge stripes, 
we need to take into account all four possible twin orientations, which are naturally present in the system due to its weak orthorhombicity. We follow the detailed description of the twinning pattern presented in Ref.~\onlinecite{braden1992characterization} (note that the tetragonal notation in the cited reference corresponds exactly to our orthorhombic cell).
This twinning structure complicates the interpretation of the data significantly. We present a detailed discussion based on the electronic stripe structure combined with the crystal twinning and illustrate the different contributions in Fig.~\ref{fig:pattern}. The typical twinning pattern of peaks along the measured $(0,k,0)$ direction is depicted in Fig.~\ref{fig:pattern}(b), where the orthorhombicity is exaggerated for clarity. The spin direction is drawn along the orthorhombic $b$-axis of the local coordinate system for each of the four twin peaks. 

As mentioned in the Introduction, the presence of magnetic stripes, which are in general incommensurate with respect to the crystal structure, is observed in neutron scattering experiments as a splitting of the antiferromagnetic $(0 1 0)$ reflection into four peaks. Signals along one diagonal, {\em e.g.}\ ${\bf Q}=(\delta,1+\delta,0)$ and ${\bf Q}=(-\delta,1-\delta,0)$, belong to charge stripe patterns along the  tetragonal $a_\textrm{T}$-axis, while the signals along the other diagonal arise due to stripe formation in the perpendicular direction, {\em i.e.}\ the $b_\textrm{T}$-axis. 
The orientation of the charge stripe patterns, either along the tetragonal $a_\textrm{T}$ or $b_\textrm{T}$-axis, are illustrated in Fig.~\ref{fig:pattern}(c).  
The commensurate structure shown here is the special case of a peak splitting of $\delta=1/8$.

In the most general case, we expect the twinning structure to be present at all four peaks in the incommensurate peak quartet, as suggested by the color coded symbols in Fig.~\ref{fig:pattern}(a). This figure furthermore provides an illustration, through the two types of symbols, of the charge stripe orientation that underlies each peak structure. Charge stripes along the tetragonal $a_\textrm{T}$-axis are depicted by triangles, while charge stripes along the tetragonal $b_\textrm{T}$-axis are shown by circles with reference to  Fig.~\ref{fig:pattern}(c). Because each twin domain is composed of a pair of twins with interchanged $a_\textrm{O}$ and $b_\textrm{O}$ axes, we obtain two sets of twins with axes oriented in the same direction, namely red/blue and yellow/green. For the green/yellow domains, stripes along the $a_\textrm{T}$-axis (charge A) will give incommensurate magnetic peaks at positions ${\bf Q}=(\delta,1+\delta)$ and ${\bf Q}=(-\delta,1-\delta)$, while stripes along the $b_\textrm{T}$-axis (charge B) will cause incommensurate magnetic peaks along the other diagonal, i.e. at ${\bf Q}=(-\delta,1+\delta)$ and ${\bf Q}=(\delta,1-\delta)$. The situation is circumvent for the red/blue domains, because the $a_\textrm{T}$ and $b_\textrm{T}$ axes are interchanged in these twins compared to the green/yellow twin domains, as illustrated in Fig.~\ref{fig:pattern}(b). 

In this way, the underlying charge structure for each twin contribution to the incommensurate signal is shown as circles and triangles in Fig.~\ref{fig:pattern}(a) with the color code as defined in Fig.~\ref{fig:pattern}(b).
The spin directions for each peak is depicted by arrows, showing the direction of the orthorhombic $b$-axis of each domain.

\begin{table}[b!]
\centering
\renewcommand{\arraystretch}{1.9}
\begin{tabular}{|c|c|c||c|c|}
  \hline
 \begin{tabular}{@{}c@{}}Neutron spin \\[-8pt] direction\end{tabular} & \begin{tabular}{@{}c@{}}Magnetic \\[-8pt] scattering\end{tabular} & Area & Isotropic & Transverse  \\ \hline
 $\frac{S \parallel x}{S \parallel y}$ & $\frac{M_{yy}+M_{zz}}{M_{zz}}$ &  $2.2 \pm 0.7$ & $2/1$ & $3/2$\\  \hline
 $\frac{S \parallel x}{S \parallel z}$ & $\frac{M_{yy}+M_{zz}}{M_{yy}}$ &  $2.5 \pm 0.8$ & $2/1$ & $3/1$\\  \hline
 $\frac{S \parallel y}{S \parallel z}$ & $\frac{M_{zz}}{M_{yy}}$        &  $1.1 \pm 0.4$ & $1/1$ & $2/1$\\ \hline
\end{tabular}
  \caption{Comparison of the fitted intensity of the inelastic polarized neutron data. The ratio of intensity in the different spin channels is shown in terms of peak area. The values are obtained by adding the fitted parameters of the 2 peaks. For the fits, the peaks position and widths are fixed to the values obtained by fitting all the data (from all 3 channels) combined. The last two columns show the expected signal ratios in the case of isotropic fluctuations and purely transverse fluctuations.}
  \label{table:amplitude ratios}
  \end{table}
 \renewcommand{\arraystretch}{1}
In the experiment, we scanned over peaks belonging to both types of charge stripes by scanning through ${\bf Q}=(-\delta,1+\delta)$ and ${\bf Q}=(\delta,1+\delta)$. The scanning direction is shown in the inset of Fig.~\ref{fig:polarization}(b). Since we only register signals where the scattering vector is perpendicular to the spins, we primarily pick up  intensity due to the red/blue domains and only negligible weight from the yellow/green domains.
Thus, we compare the relative strength of the charge types A and B of the red/blue twins and we can decide whether the system displays charge stripes along only one of the tetragonal axes or along both directions. 

As shown in Fig.~\ref{fig:polarization}(b), we find that the peaks at ${\bf Q}=(-\delta,1+\delta)$ and ${\bf Q}=(\delta,1+\delta)$ have similar amplitudes.  
From the equal signal amplitudes at both incommensurate positions, we conclude that charge stripes form along both the $a_\textrm{T}$- and $b_\textrm{T}$-direction and are equally present within the sample. 
In the event that charge stripes had a preferred direction (parallel or perpendicular to the $b_\textrm{T}$-axis), only one of the two IC peaks would have been visible. 
Our finding that the system displays charge stripes along both tetragonal axes is  similar to the observation in LBCO with $x=\frac{1}{8}$.~\cite{agterberg2020physics} However, our data does not provide information about a possible orthogonal arrangement of stripes in adjacent planes, as discussed in the case of LBCO \cite{Hucker2011a} and most likely, phase sensitive measurements are required to pursue this further.

We now turn to the inelastic polarized data of Fig.~\ref{fig:polarization}(a). As opposed to the static signal, which shows that spins are purely in-plane, the inelastic scattering signal at $\hbar \omega=1.5$~meV shows a clear out-of-plane spin component. First, we consider the expected outcome of the signal ratios in the case of isotropic spin fluctuations, where fluctuations along all three spin directions are equal in magnitude, {\em i.e.}\ where the transverse and longitudinal fluctuations are equally strong. Then we would expect to see the same signal strength in the spin channels $S \parallel y$ and $S \parallel z$ and double intensity strength in the spin channel $S \parallel x$, i.e.\ a signal ratio of $2/1/1$ for the spin channels $x/y/z$.
At the opposite end, we consider isotropic transverse fluctuations with longitudinal fluctuations being negligible,   
%perpendicular to spin moments ordered along the orthorhombic $b$-axis 
i.e. transverse fluctuations of equal strength in all directions perpendicular to the spin direction, but no fluctuations in the direction of the ordered moment. In this case,
we would expect a signal ratio of $3/2/1$ taking into account the twinning structure. To distinguish between these two limiting cases, we calculate the ratio between the peak areas of the three spin channels with the results shown in the third column of Table ~\ref{table:amplitude ratios}. These results are compared to the expected ratios in the case of isotropic and purely transverse fluctuations, see fourth and fifth columns of 
Table ~\ref{table:amplitude ratios}. This provides guidance to the dominating nature of the incommensurate spin fluctuations at $1.5$~meV.

The data indicates that the spin fluctuations are isotropic, in particular because the signals with polarization along the $y$- and $z$-directions are almost equal. However, we have only taken into account the extreme cases. It should be noted that intermediate cases could lead to similar signal ratios. For example, anisotropic transverse fluctuations with a more pronounced in-plane component (in the $x$-$y$ plane) than out-of-plane component could also result in a a signal ratio of $2/1/1$ for the spin channels $x/y/z$ which would be compatible with the obtained signal ratios. A detailed study of fluctuations, with measurements also along the $l$-direction in reciprocal space has the potential to shed light onto this matter. Due to long counting times, imposed by the polarization set-up and the intrinsic weak magnetic inelastic signal, combined with the necessity to realign the sample, we have not been able to further pursue this idea.  

Another perspective on the relation between static order and low-energy spin fluctuations in the very same sample was acquired recently by some of us in Ref.~\onlinecite{Jacobsen2018}. A difference in incommensurability between the static signal and low-energy spin fluctuations led us to conclude that magnetic order and fluctuations likely originate from separate spatial domains within the crystal. If the nature of the low-energy spin fluctuations had been primarily transverse, i.e.\ spin-wave-like Goldstone modes of the static spin order, this would have contradicted the findings of Ref.~\onlinecite{Jacobsen2018}. The observation that the $1.5$~meV fluctuations are more likely of isotropic character supports the interpretation in Ref.~\onlinecite{Jacobsen2018} and re-affirms our hypothesis that the low-energy fluctuations reside in different parts of the crystal than the static order. These regions could very well be where superconductivity is present, as we will discuss in the next section.

\subsection{Phase separation and intertwined orders}
Figure ~\ref{fig:susceptibility and temp dependence elastic} demonstrates that the sample exhibits similar critical temperatures for superconductivity and magnetic order, i.e. $T_{\rm c} \simeq$ $T_\textnormal{N}$. This finding is supported 
by results from local probe nuclear magnetic resonance (NMR) measurements\cite{ImaiNMR2018} and was also found by neutron scattering\cite{Lee1999a, lee2004neutron} in LCOO samples with the same critical temperature as the one used in this study.
While this could be interpreted as a microscopic coexistence of magnetic and superconducting order,~\cite{lee2004neutron} there is also the possibility of a microscopic phase separation of the crystal into different domains where each contains only one type of order, either magnetic or superconducting, both phases with comparable free energy.~\cite{Savici2002}

Evidence of phase separation was found in the local probe NMR and $\mu$SR studies of similar systems,~\cite{ImaiNMR2018,Savici2002,Mohottala2006,Udby2013} and in a neutron scattering study of this very same sample.~\cite{Jacobsen2018} Taking this into account, some of us have previously advocated for a phase separation in which one part of the crystal shows static magnetic order and associated Goldstone modes, while other parts of the crystal display low-energy fluctuations without static order at a slightly different incommensurability. 
We speculate that the Goldstone modes associated with the static magnetic order are too weak to be detected with neutrons when superimposed on the signal from the low-energy fluctuations.

Another proposal of phase separation in the cuprate family was put forward in the case of underdoped LSCO\cite{Kofu09}. There, the electronic structure of the sample was interpreted as divided into two phases; 
magnetic order with low-energy fluctuations ($< 4$~meV) in some parts of the crystal and gapped higher energy fluctuations and superconductivity in other parts of the crystal. 
Evidence for this phase separation was: a) a dip in the energy spectrum at $\sim 4$~meV and b) a small change of linewidth of the signal at this energy. 
We do not observe neither a dip nor an abrupt change of linewidth as a function of energy. However, we note that such effects are very subtle and hard to detect, and we therefore hesitate to draw definite conclusions based on our observations.

In the large number of studies performed on the La$_{2-x}$M$_x$CuO$_{4+y}$ family of superconductors,
we observe two general types of magnetic behaviour in the superconducting state: 1) compounds that do not exhibit magnetic order and have gapped low-energy spin excitations below $T_{\rm c}$ and 2) compounds where magnetic order is present concomitantly with a continuous spectrum of spin fluctuations, albeit with reduced strength at low energies. Our stage-4 LCOO sample belongs to the latter category, exhibiting both static stripe order as shown in Fig.~\ref{fig:polarization}(b) and ungapped low-energy fluctuations displayed in Fig.~\ref{fig:raw data}(b). We note that fluctuations appear insensitive to the onset of superconductivity. In fact there is a strengthening of the low-energy spin susceptibility at the lowest temperatures. The suppression of the very low-energy signal ($\hbar\omega < 4$~meV) with increasing temperature is likely a consequence of decreased spin coherence. Figure~\ref{fig:susceptibility and temp dependence elastic}(c) points to a complete destruction of spin coherence at $T\sim 70$~K. 
Our sample is therefore different from optimally doped LSCO, which does not show static order and in which low-energy spin fluctuations are gapped in the superconducting state and furthermore persist up to $350$~K.~\cite{aeppli1997nearly} The latter feature is depicted by the grey data points in Fig.~\ref{fig:raw data}(b) which are reproduced from Refs.~\onlinecite{Lake01,Lake2002}. At doping $x\simeq 0.16$, LSCO exhibits a spin gap of roughly $\Delta_{0.16} \sim 6$-$7$~meV. This sample has a similar optimal value of $T_{\rm c} \simeq 39 $~K as found in LCOO. In the event that LCOO had a (magnetically) similar superconducting phase as in optimally doped LSCO, this would show up as an incomplete spin gap around $\Delta_{0.16}$ in our measurements. This is clearly not the case.

We can think of three scenarios that would be compatible with the lack of spin gap: 1) if superconductivity arises in yet other parts of the crystal, where there are no low-energy spin fluctuations of $\hbar \omega <7$~meV, 2) an increase in the low-energy fluctuations in the non-superconducting part of the sample exactly matches a decrease of the fluctuations in the superconducting regions or 3) superconductivity coexists with and does not compete with low-energy magnetic fluctuations. The two first scenarios seem unlikely, since 1) would imply regions with a very different low-energy electronic behavior not observed in any other superconductors to date
and 2) would seem like an improbable coincidence.
Thus, we are left with the third scenario, which points to a coexistence of low-energy fluctuations and superconductivity in LCOO. 

In underdoped LBCO, a lack of spin gap has also been observed and interpreted as evidence of a PDW type of electronic structure with intertwined modulated superconducting order and spin stripes~\cite{Xu2014,christensen2016magnetic} and points to a common origin of the magnetic order and fluctuations.
In our sample, we interpret the absence of any significant decrease in $\chi''$ for $T <$ $T_{\rm c}$ at energies $\hbar \omega \leq 10 $~meV, as an indication that superconductivity in LCOO microscopically coexists with the low-energy magnetic fluctuations. However, because of additional evidence that the low-energy fluctuations do not occur in the same spatial regions as the static magnetic order,~\cite{Jacobsen2018} this proposed coexistence phase is different from the usual PDW phase: Only low-energy fluctuations are intertwined with superconductivity, while magnetic order appears as a separate, probably competing, phase. In addition, by presenting our data on an absolute scale, we find that the magnetic spectral weight of low-energy fluctuations in LCOO is approximately one order of magnitude lower than that of LBCO or the parent compound, LCO (see the Supplementary Material). In LBCO, the strong magnetic response was taken as indication that magnetic order and fluctuations coexist locally with superconductivity in the entire bulk of the sample. The reduction of magnetic spectral weight in LCOO found here might be due to the fact that a fraction of the sample shows AFM order. The corresponding magnetic signatures of the AFM order are not included in our analysis, and this would lead to an apparent reduction in the spin susceptibility when presented in absolute units.
%This evidence supports the phase separation scenario in which small regions of the sample exhibit fluctuations without magnetic order, in contrast to LBCO where the strong magnetic response was taken as indication that magnetic (order and fluctuations) and superconducting phase coexist locally in the entire bulk of the sample. 

The temperature dependence of the magnetic susceptibility of LSCO with $x=1/8$ doping~\cite{Roemer13} is remarkably similar to our findings in LCOO. This might indicate a similar coexistence of low-energy fluctuations and superconductivity in that system as well. Note that an additional similarity of LCOO and LSCO $x=12$ \% is the curious insensitivity to a magnetic field of the low-energy fluctuations, which could provide further support for a coexistence phase.

%Finally, we point to the fact that Wakimoto {\em et al.} earlier found low-energy spin excitations of overdoped LSCO samples ($0.25 \leq x \leq 0.28 $) to be gapless in the superconducting state and insusceptible to a temperature increase~\cite{wakimoto2004direct} and in their work pointed to the possibility of a cooperative relation between low-energy fluctuations and superconductivity. Even though overdoped LSCO is singled out by not showing static magnetic stripes, the lack of the spin gap in this system is curiously similar to our observations in LCOO, as well as the observations in LBCO and underdoped LSCO.
%~\cite{Xu2014,Roemer13,Kofu09}. 
%, the  and could point to a direct relation between the spin fluctuations and superconductivity in overdoped LSCO as well as in LCOO.

To further elucidate the nature of the superconducting state, it would be interesting to investigate the presence or absence of a magnetic resonance peak by inelastic neutron measurements of high energy spin excitations. One of the fingerprints of the PDW is the absence of the magnetic resonance peak~\cite{Xu2014,christensen2016magnetic} otherwise observed, in standard $d$-wave SC, at E$_{\rm cross} \approx 40$-$50$~meV.~\cite{NBC_hourglass_2004,vignolle2007two, fujita2012progress}

\subsection{Magnetic field effect}
\label{discussion field effect}

A uniform $d$-wave superconductor is expected to show magnetic field-enhancement of the magnetic spectral weight at low energies since the destruction of superconductivity in the vortex cores paves the way for magnetic order.~\cite{Brian2011} This is opposite to our observations in LCOO, see Fig.~\ref{fig:susceptibility field depednence}(a), where we instead of a field-enhancement find a small suppression of the spectral weight at low energies in the range $\hbar \omega=0.7$-$1.5 $~meV. In optimally doped LSCO samples, the sub-gap field-induced states are thought to originate from a separate electronic phase most likely prevailing in the vicinity of the vortices.~\cite{Lake01,chang2008tuning} This explanation is similar to the above-mentioned phase separation scenario proposed by Kofu {\em et al.}~\cite{Kofu09} where low-energy fluctuations, in underdoped LSCO samples, are thought to originate from a phase different from the superconducting one. In our sample, on the contrary, we argue that the low-energy fluctuations reside within the
superconducting regions. 
Because of this, we do not expect a significant effect induced by the appearance of a vortex lattice in the superconducting regions. This is in agreement with our observations of a very weak suppression of the low energy signal in a $12$~T applied magnetic field.

On the other hand, in terms of elastic stripes, in opposition to the lack of field effect in our sample (see Fig.~\ref{fig:susceptibility field depednence}(c)), Khaykovich {\em et al.}~\cite{khaykovich2002enhancement, khaykovich2003effect} and Lee {\em et al.}~\cite{lee2004neutron} have previously reported a significant field enhancement of the elastic signal, comparable to the effect of electronic disorder induced by quenched cooling, in both stage-4 and stage-6 LCOO samples. %This is clearly not the case in our sample where an applied magnetic field produces no detectable effect (see Fig.~\ref{fig:susceptibility field depednence}(c)). 
Taking into account the high quality of our crystal, which exhibits a sharp single transition to superconductivity at $42$~K (Fig.~\ref{fig:susceptibility and temp dependence elastic}(a)), we can think of two possible explanations for this discrepancy: 1) the slow cooling procedure we used in our experiments (with a cooling rate of $1$~K/min) could differ from that used previously and result in a different arrangement of the excess oxygen, leading to a different electronic configuration at base temperature; 2) subtle differences in the oxygenation procedure could have created samples with different ground states containing various ratios of superconducting and magnetically ordered phases.  

With respect to the first explanation, the exact cooling rate used during the field experiments is not stated in the literature. The only value we can compare against is the quench cooling rate of $\sim 2.8$~K/min~\cite{lee2004neutron} that creates a significant structural distortion which in turn induces an enhancement of the magnetic order. Only if our sample were to be in a highly disordered phase, following the cooling procedure, the effect of an applied magnetic field would be as insignificant as the one we have measured. This is highly unlikely, since our much lower cooling rate of $1$~K/min is expected to give rise to an ordered oxygen lattice at low temperatures.

The second, and more likely, scenario is based on observations by Chang {\em et al.},~\cite{chang2008tuning} demonstrating that the effect of an applied magnetic field is to enhance the magnetic order in LSCO samples towards a common plateau level, which corresponds to the ground state of systems with $1/8$ doping value. In this case, the initial fraction of the magnetic ordered phase in the sample dictates the magnitude of the impact of an applied magnetic field. Thus, the lack of a field effect which we have measured, indicates that our sample, in contrast with the ones used by Khaykovich {\em et al.} and Lee {\em et al.}, exhibits a fully developed magnetic phase already in zero applied magnetic field. We note that this does not prohibit the existence of a superconducting phase with similarly high $T_{\rm c}$ compared to samples with magnetic phases of different magnitudes.\\

\section{Conclusion}

We have presented an extensive neutron scattering study which characterizes the magnetic order and fluctuations in the stage-$4$ oxygen doped La$_2$CuO$_{4+\delta}$ superconductor. 

The lack of a spin gap 
%or magnetic field effect 
in the low-energy magnetic spectrum is interpreted as an indication that superconductivity and spin fluctuations co-exist microscopically in an intertwined phase with no sign of competition. Notably this happens in a single layer cuprate superconductor that exhibits a clean superconducting transition with one of the highest critical temperatures of $T_{\rm c} = 42$~K.

 On the other hand, we consider magnetic order to be a separate electronic phase competing with superconductivity, in agreement with previous interpretations from the literature.~\cite{julien2003magnetic} The lack of a magnetic field effect on the elastic stripe signal, 
shows that our sample contains a fully developed magnetic phase. 
This observation contrasts previous reports of a magnetic field enhancement,~\cite{khaykovich2002enhancement, khaykovich2003effect, lee2004neutron}, and is possibly explained by discrepancies in the initial magnetic volume fraction of crystals which have been doped under different oxygenation procedures. 

The $xyz$ polarization analysis revealed the presence of equal fractions of spin stripe modulations %induced by charge stripes 
oriented along either of the tetragonal $a_\textrm{T}$-axis or $b_\textrm{T}$-axis. 

Corroborated with a study previously published by some of us,~\cite{Jacobsen2018} our data supports the hypothesis that the magnetic fluctuations in LCOO are likely isotropic in nature, but further work is needed to confirm this. 

In conclusion, we propose a picture of electronic phase separation in LCOO samples into two competing phases, where one exhibits magnetic stripe order while the other hosts superconductivity intertwined with low-energy spin fluctuations.

\section{Acknowledgements}
We would like to give a special thanks to Niels Hessel Andersen for his work within this field of science throughout his long career, and particularly thank him for his contributions to this project. He will be missed, both as a colleague and as a friend. 

We are grateful for the access to neutron beamtime at the ILL and HZB neutron facilities. This project was supported by the Danish Council for Independent Research through DANSCATT. ATR acknowledges support from the Carlsberg Foundation. HJ acknowledges funding from the EU Horizon 2020 program under the Marie Sklodowska-Curie grant agreement No 701647. AET was supported through the ILL Ph.D. program. SHD acknowledges support from Nordforsk under NNSP.

% \bibliography{LSCOOinelastic}
\bibliography{LSCOOinelastic_corrected}

\clearpage
\appendix
\twocolumngrid
\section{Supplementary Material}

\subsection{Magnetic susceptibility}

\renewcommand{\arraystretch}{1.5}
\begin{table}[b]
\centering
\begin{tabular}{|c|c|c|}
  \hline
 Element & Atom's mass (m) [g]& No. of atoms/unit cell (n)  \\ \hline
  La & 2.31 $\cdot$ 10$^{-22}$ & 8   \\ \hline
  Cu & 1.05 $\cdot$ 10$^{-22}$ & 4    \\ \hline
  O & 2.66  $\cdot$ 10$^{-23}$ & 16.32 \\ \hline
\end{tabular}
  \caption{The mass of one atom was calculated as the ratio between the atomic mass and Avogadro's number. The number of atoms is presented for a conventional unit cell of LCO+O with oxygen doping $\delta=0.08$.}
  \label{table:atoms}
  \end{table}
  
In order to accurately determine the magnetic (volume) susceptibility of the sample in SI units, from the magnetic moment measurements, the following formula was applied:

\begin{equation}
\chi = 4\pi\,\frac{\mu}{V \cdot H_{internal}},
\end{equation}
 where $\mu$ is the measured magnetic moment in electromagnetic units (emu), $V$ is the sample's volume in cm$^3$ and $H_{internal}$ is the strength of the internal magnetic field. It should be noted that the internal magnetic field differs from the applied one ($H_{applied}$) by a demagnetizing factor ($N$): 
 
 \begin{equation}
 H_{internal} =  \frac{1}{1-N}H_{applied},
 \end{equation}
which is valid only in the Meissner state, i.e. for very small applied magnetic fields. Our measurements were performed under $H_{applied} = 6.4$~mT.

 If we treat our cubic samples as spheres, in which case the demagnetizing factor has the value N = 1/3, we obtain:
  \begin{equation}
 H_{internal} = 1.5 \times H_{applied},
 \end{equation}
 
For an accurate measure of the sample's volume, the density ($\rho$) of LCOO is used in the calculations as follows:
\begin{equation}
\label{eq:density}
\rho = \frac{m_{UnitCell}}{V_{UnitCell}}.
\end{equation}
where $V_{UnitCell}$ is the unit cell volume calculated with the lattice parameters taken from Ref.~\onlinecite{radaelli1993structure} for oxygen doping $\delta=0.08$ at a temperature of $10$~K:

\begin{equation}
\begin{split}
  V_{UnitCell} & = a \times b \times c = 5.33\,\textnormal{\AA} \times 5.39\,\textnormal{\AA} \times 13.16\,\textnormal{\AA} \\
& = 378.07 \, \textnormal{\AA}^3 = 378.07 \cdot 10^{-24} \,\textnormal{cm}^3.  
\end{split}
\end{equation}

The number of atoms per unit cell and the atomic masses of the 3 components of LCOO (shown in Table~\ref{table:atoms}): are used to determine the mass of the unit cell:

\begin{equation}
\begin{split}
 m_{UnitCell} & = m_{La}\cdot n_{La} + m_{Sr}\cdot n_{Sr} + m_{Cu}\cdot n_{Cu}+ m_{O}\cdot n_{O} \\
& = 2.70 \cdot 10^{-21} \textnormal{g}.   
\end{split}
\end{equation}

Finally, following equation~\ref{eq:density}, the density of LCOO with doping $x=0.08$ is  \textbf{$\boldsymbol{\rho = 7.14}$ g/cm$\boldsymbol{^3}$}. This value is afterwards used as the division factor of the mass in order to obtain an accurate estimation of the sample's volume.

The calculated susceptibility is then plotted against the temperature as it can be observed in Figure~\ref{fig:susceptibility and temp dependence elastic} in the main text. In an ideal case, the graph will have the shape of a sigmoid function constrained in between 0 (corresponding to the antiferromagnetic state above $T_{\rm c}$) and -1 (corresponding to the perfect diamagnetic response of the material below $T_{\rm c}$). 

\subsection{Absolute normalization of magnetic cross section}

\newcommand{\qqq}{\mathbf{q}}
\newcommand{\QQQ}{\mathbf{Q}}
\newcommand{\GGG}{\mathbf{G}}

The normalization process of the magnetic cross section follows closely the procedure explained by Xu {\em et al.} in Ref.~\onlinecite{xu2013absolute}. Here we exemplify how we have applied it on our data, in particular the data collected on IN12. We note that special attention should be paid to the units used. 

The dynamic susceptibility is defined as:

\begin{equation}
\label{eq chi}
    \chi'' (\QQQ,\omega) = \frac{\pi}{2} \mu_B^2 (1-e^{-\hbar \omega/k_B T})\frac{13.77(b^{-1}) \tilde{I}(\QQQ,E)}{ |f(\QQQ)|^2 e^{-2W} N k_f R_0}.
\end{equation}

We will now illustrate the calculation of each of the terms for the IN12 data.

\begin{itemize}
    \item $N k_f R_0$ - the resolution volume
\end{itemize}

The resolution volume is obtained from the sample phonon scattering measured as a constant energy scan:

\begin{equation}
\hspace*{-0.2 cm}
    Nk_fR_0 = \frac{\int\tilde{I}(\QQQ,E) d\qqq }{e^{-2W} |F_N(\GGG)|^2 \cos^2(\beta) \frac{m}{M} \frac{(\hbar \QQQ)^2}{2m} \frac{n}{\hbar \omega} \frac{1}{d\omega/dq} },
\end{equation}
where $\int\tilde{I}(\QQQ,E) d\qqq$ is the average integrated intensity of the two phonon branches. 
$e^{-2W}$ is the Debye-Waller factor and $\beta$ is the angle between $\QQQ$ and the polarization of the phonon. Both $e^{-2W}$ and $\cos^2(\beta)$ are assumed 1.
$F_N(\GGG)$ is the structure factor of the phonon, it can easily be obtained from VESTA~\cite{momma2008vesta} and should be used in units of barns.
$m$ is the neutron mass and $M$ is the is the mass of all the atoms in the unit cell (see Table~\ref{table:atoms}).
$n=1/(1-e^{-E/k_BT})$ is the Bose factor of the phonon, where $E$ is the energy at which the phonon was measured, $k_B$ is the Boltzmann constant and $T$ is the temperature at which the phonon was measured. $d\omega/dq$ is the phonon velocity where $\omega$ is the energy at which the phonon was measured and $q$ is obtained as half the distance between the two phonon branches in reciprocal lattice units.
The term $\frac{(\hbar \QQQ)^2}{2m}$ should be calculated in units of meV.%, $\hbar$ should be used in both Js $\equiv$ kg m$^2$/s and meVs units, $\QQQ$ should be used in units of m$^{-1}$ and $m$ in units of kg.

In the case of our acoustic phonon measured on IN12 at $(2,0,0)$, $3$~meV and $280$~K, we obtain the numerical values shown in Table~\ref{tab:phonon_values}. 

\renewcommand{\arraystretch}{2}
\begin{table}[h!]
\resizebox{0.5\textwidth}{!}{%
    % \centering
    \begin{tabular}{|c|c|c|c|c|c|c|cc}
    \hline
       \textbf{Term}  & $F_N(\GGG)$ & $m/M$ & $\frac{(\hbar \QQQ)^2}{2m}$ & $\frac{n_q}{\hbar\omega}$ & $d\omega/dq$ & $N k_f R_0$\\ \hline
       \textbf{Unit} & barns & - & meV & meV$^{-1}$ & meV/ r.l.u. & meV/ barns
       \\ \hline
       \textbf{Value} & 91 & $\frac{1}{1617}$ & 11.57 & 2.85 & 35.76 & 0.0025
        \\ \hline
    \end{tabular}}
    \caption{Parameters corresponding to the acoustic phonon measured on IN12.}
    \label{tab:phonon_values}
\end{table}

\begin{itemize}
    \item $f(\QQQ)$ - the form factor
\end{itemize}

In the case of LCOO it is sufficient to use the Cu form factor since it is the only atom that is responsible for the magnetic scattering 
\begin{equation}
    f(\QQQ) =  A e^{-a (\frac{\QQQ}{4\pi})^2}+B e^{-b (\frac{\QQQ}{4\pi})^2} +C e^{-c (\frac{\QQQ}{4\pi})^2} + D,
\end{equation}
where the coefficients $A$, $a$, $B$, $b$, $C$, $c$ and $D$ can easily be found tabulated~\cite{ILLFormFactors} for Cu$^{2+}$.

%\columnbreak

The other remaining factors in Eq.~(\ref{eq chi}) are the Bose factor of the measured magnetic signal $1/(1-e^{-E/k_BT})$, the Debye-Waller factor $e^{-2W}$ assumed $1$, the Bohr magneton $\mu_B$ and the amplitude of one of the incommensurate magnetic peaks $\tilde{I}(\QQQ,E)$.

Furthermore, in order to obtain the $\QQQ$-integrated local susceptibility ($\chi''(\omega)$), as presented in Ref.~\onlinecite{Xu2014} for an underdoped LBCO sample, an additional integration over all the peaks in the Brillouin Zone (BZ) is needed:

\begin{equation}
\label{eq chi avg}
    \chi'' (\omega) = \frac{2\pi\sigma_h \sigma_k n_{peaks}}{a^*b^*}\chi'' (\QQQ,\omega),
\end{equation}
where $\sigma_h$ and $\sigma_k$ are the widths of the gaussian fits to the peaks in $h$- and $k$-directions, $n_{peaks}$ is the number of peaks in the BZ and $a^*, b^*$ are the reciprocal lattice parameters. In our orthorhombic notation there are 2 peaks in the BZ and, since we have only performed scans along one of the in-plane reciprocal space directions, we assume that the incommensurate peaks are symmetrical ($\sigma_h = \sigma_k$).  

\begin{figure}[h!]
\includegraphics[width=0.45\textwidth]{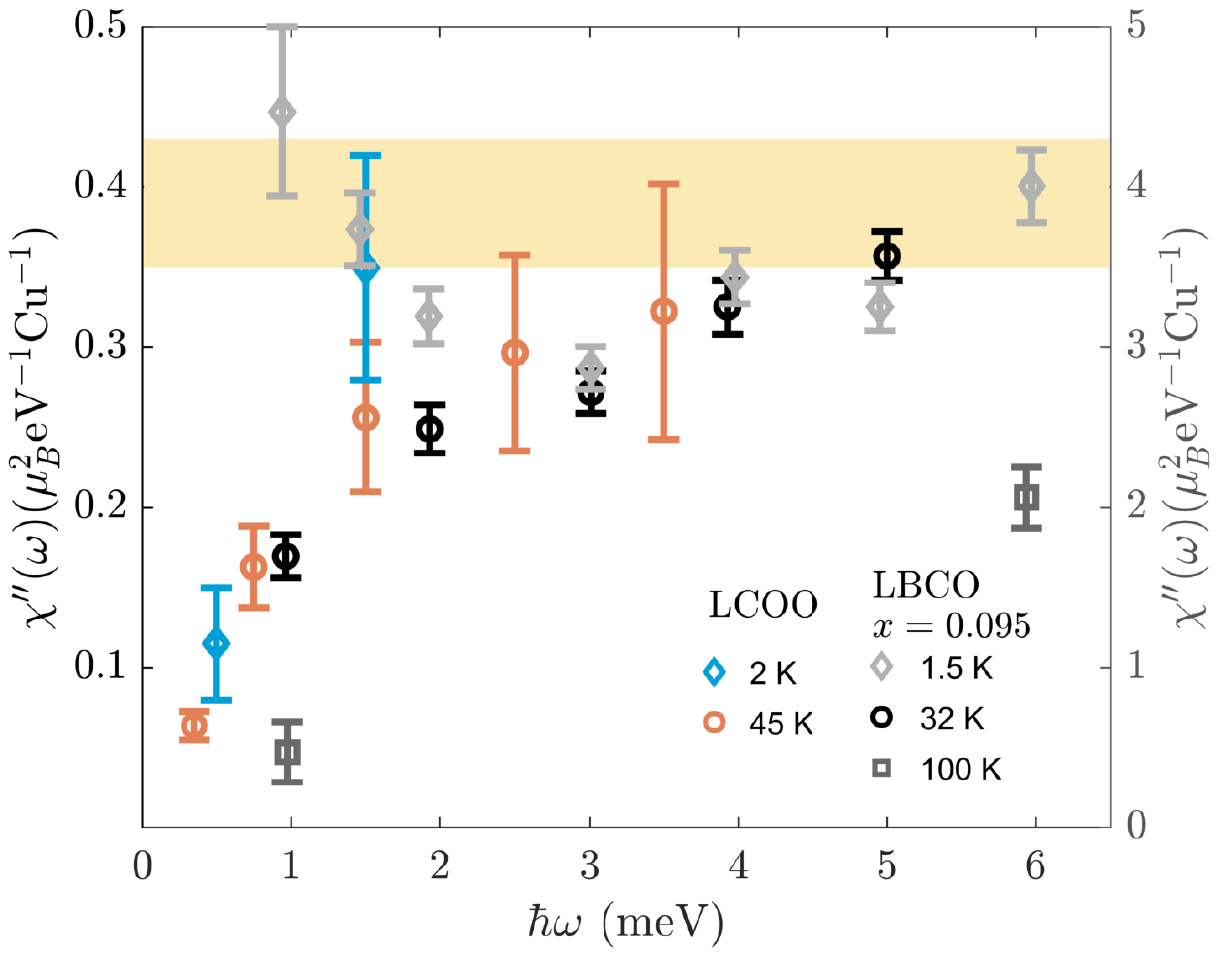}
\caption{Dynamic susceptibility $\chi''(\omega)$ measured at temperatures inside ($2$~K) and outside ($45$~K) the superconducting dome. All data has been measured at Thales (Th1) with constant $k_\textnormal{f}=1.55$~\AA{}$^{-1}$. In grey and black symbols, we show the magnetic susceptibility measured inside and outside the superconducting state by Z. Xu {\em et al.}~\cite{Xu2014} on an underdoped $x=0.095$ LSBCO sample with $T_{\rm c}= 32$~K. The shaded area indicates the magnitude of the spin waves in the parent compound LCO, as presented by Z. Xu {\em et al.}~\cite{Xu2014}, and is connected to the right hand side axis.}
\label{fig:comparison LBCO Thales}
\end{figure}

We note that $\chi''(\omega)$ does not depend on the choice of unit cell (tetragonal or orthorhombic), meaning that all variations of the parameters (such as the number of Cu per unit cell or the lattice parameters) eventually cancel out. We can thus make a direct comparison, as shown in Figure~\ref{fig:comparison LBCO Thales},  between the magnetic spectral weight of our sample and the one of an underdoped LBCO crystal presented in the literature.  

For the integration we have assumed that the peak is Gaussian with the same width along $h$ and $k$. In general, the peak is expected to elliptical, with the minor and major axes not necessarily along $h$ and $k$. This assumption thus leads to a fairly large uncertainty in $\chi''(\omega)$. 

Despite these shortcomings of the absolute normalisation process, the procedure allows for a rough comparison of the magnetic spectral weight of the very low-energy fluctuations in LCOO with that of underdoped LBCO and the parent compound LCO.~\cite{Xu2014} As shown in Fig.~\ref{fig:comparison LBCO Thales}, we find that the magnetic weight in LCOO is one order of magnitude lower than in LBCO and undoped LCO. We ascribe this to the fact that some regions of our LCOO crystal show AFM order, which we do not include in the analysis.
%This fits very well with the phase separation scenario proposed here in which only parts of the sample exhibit magnetic excitations.

\end{document}